\documentclass{jfm}
\usepackage{graphicx}
\usepackage{epstopdf,epsfig}
\usepackage{natbib}
\usepackage{hyperref}
\usepackage{natbib}
\usepackage{amsmath,mathtools}
\usepackage{graphicx}
\usepackage{adjustbox}
\hypersetup{
    colorlinks = true,
    urlcolor   = blue,
    citecolor  = black, 
}
\usepackage{color}
\usepackage{bm}
\usepackage{subcaption} 
\newcommand{\RomanNumeralCaps}[1]

\def\beq{\begin{equation}}
\def\eeq{\end{equation}}
\newcommand{\ks}{k_\star}
\newcommand{\Ac}{\mathcal{A}}
\newcommand{\Bc}{\mathcal{B}}
\newcommand{\Fc}{\mathcal{F}}
\newcommand{\Sc}{\mathcal{S}}
\newcommand{\Fcs}{\mathcal{F}_{\!\star}}
\newcommand{\Acs}{\mathcal{A}_{\star}}
\newcommand{\cs}{c_\star}

\newcommand{\sigmas}{\sigma_\star}

\newcommand{\delsig}{\delta_{\sigma}}

\newcommand{\Hs}{ H_{\mathrm{s}} }
\newcommand{\hs}{ h_{\mathrm{s}} }
\newcommand{\Hss}{ H_{\mathrm{s}\star} }

\newcommand{\zetrms}{\zeta_{\mathit{rms}}}
\newcommand{\zetrmss}{\zeta_{\mathit{rms}\star}}
\newcommand{\Ds}{D_\star}

\newcommand{\Th}{\Theta}
\newcommand{\tha}{\theta}
\newcommand{\del}{\delta}

\newcommand{\defn}{\stackrel{\text{def}}{=}}

\newcommand{\rv}{r_{\mathit{v}}}
\newcommand{\ells}{\ell_{\mathit{s}}}

\newcommand{\erfc}{\mathrm{erfc}}

\def\ee{\mathrm{e}}

\def\bx{{\bm{x}}}
\def\by{{\bm{y}}}
\def\bk{{\bm{k}}}
\def\bU{\bm{U}}

\def\eps{\varepsilon}
\def\dd{\mathrm{d}}

\def\Hs{H_s}

\newcommand{\half}{\tfrac{1}{2}}

\definecolor{HW}{RGB}{137,0,225}

\newcommand{\LL}{\left}
\newcommand{\RR}{\right}
\def\hs{h_s}

\newcommand\change[1]{#1}

\title{Scattering of swell by currents}

\author{Han Wang\aff{1} \corresp{\email{hannnwangus@gmail.com}}, 
Ana~B. Villas B\^oas\aff{2}, 
William R. Young\aff{3}
  \and
  Jacques Vanneste\aff{1}}

\affiliation{\aff{1}School of Mathematics and Maxwell Institute for Mathematical Sciences, University of Edinburgh, EH9 3FD, UK
 \aff{2}Department of Geophysics, Colorado School of Mines, Golden CO 80401, USA
 \aff{3} Scripps Institution of Oceanography, University of California at San Diego, La Jolla CA 92093-0213, USA
 }

\begin{document}
\maketitle

\begin{abstract}
The refraction of surface gravity waves by currents leads to spatial modulations in the wave field and, in particular,  in the significant wave height. 
We examine this phenomenon in the case of waves scattered by a localised current feature, assuming (i) the smallness of the ratio between current velocity and wave group speed, and (ii) a swell-like, highly directional wave spectrum. 

We apply matched asymptotics to the equation governing the conservation of wave action in the four-dimensional position--wavenumber space. The resulting explicit formulas show that the modulations in wave action and significant wave height past the localised current are controlled by the vorticity of the current integrated along the primary direction of the swell. 

We assess the asymptotic predictions against numerical simulations using WAVEWATCH III for a Gaussian vortex. We also consider vortex dipoles to demonstrate the possibility of `vortex cloaking' whereby certain currents have (asymptotically) no impact on the significant wave height.  We discuss the role of the ratio of the two small parameters characterising assumptions (i) and (ii) above and show that caustics are only significant for unrealistically large values of this ratio, corresponding to unrealistically narrow directional spectra.

\end{abstract}

%
%

\section{Introduction}
\label{sec:intro}
Surface gravity waves (SGWs) play a key role in the exchanges of energy, momentum and gases between the ocean and the atmosphere \citep{villasboaspizzo2021}. 
SGWs are forced by the wind and modulated by ocean currents through transport and refraction.  Over the past few decades, several studies have explored the effects of ocean currents on SGWs. Early theoretical work focusses on the formation of freak waves and identifies refraction as a possible mechanism for the generation of large amplitude waves \citep{WhiteFornberg1998, Heller2008, dysthe2008oceanic}.

Recent studies examine how meso- and submesoscale ocean variability, such as fronts, filaments and vortices, induces a corresponding variability in wave amplitudes \citep{Ardhuin2017, Romero2017, Romero2020, VillasBoas2020,VPL2022}. 
These studies often characterise the wave amplitudes using  the significant wave height $\Hs$, defined as four times the standard deviation of the surface displacement.
They find that wave--current interactions at horizontal scales ranging from 10 to 200 km drive spatial gradients of $\Hs$ at similar scales. This indicates  that air--sea fluxes might have spatial variability on these relatively small spatial scales.

One common approach to studying wave--current interactions is the use of ray tracing, often in its simplest form in which  the kinematics of SGWs is tracked by solving the ray equations and ray density is used as a proxy for wave amplitude \citep[e.g.,][]{Kenyon1971, mapp1985WaveRefraction,Quilfen2019}. 
While this simple form of ray tracing is a valuable tool for understanding wave refraction, it does not provide an accurate quantification of changes in wave amplitude, in particular changes in $\Hs$. This quantification requires to solve the conservation equation for the density of wave action in the four-dimensional position--wavenumber phase space. This is challenging especially for the wave spectra of realistic sea states,
distributed in both wavenumber and direction, instead of the pure plane waves that are often considered \citep[see][however]{Heller2008}.  
It is possible to solve the action equation numerically, albeit at great computational cost, either by discretising  the phase space or by sampling its full four-dimensionality with a large ensemble of rays.  

This paper proposes a complementary approach. It develops an asymptotic solution of the wave action equation, leading to explicit formulas for the changes in action and $\Hs$ induced by localised currents. Motivated by their ubiquity in the ocean, we focus on swell, that is, SGWs characterised by a spectrum that is narrow banded in both frequency (equivalently, wavenumber) and direction. We exploit the smallness of two parameters reflecting the narrowness of the spectrum and the weakness of the current relative to the wave speed. \change{We approximate the wave action equation to leading order and solve it in closed form by integration along its characteristics (the approximate ray equations) by inspection.}
The formulas we obtain show that the changes in action and $\Hs$ depend on the currents through a `deflection function' $\Delta$ given by the integral of the vorticity along the primary direction of wave propagation. We apply these formulas to simple flows -- vortices and dipoles -- and compare their predictions with the results of full integrations of the action conservation equation by a numerical wave model.

We formulate the problem, relate action and $\Hs$, and introduce a model spectrum for swell in  \S\ref{sec:formulation}. We detail our scaling assumptions and carry out the (matched) asymptotics treatment of the wave action equation in \S\ref{scatterSec}. We compare asymptotic and numerical results for vortices and dipoles in \S\ref{simpleFlows}. For vortices, we consider four different parameter combinations representative of  ocean swell. We consider dipoles with axis along and perpendicular to the direction of the swell to demonstrate the possibility of a vanishing deflection function $\Delta$, leading to asymptotically negligible changes in $\Hs$, a phenomenon we refer to as `vortex cloaking'.  In \S\ref{sec:limits} we explore two limiting regimes of scattering: a linear regime, corresponding to weak currents and/or swell with relatively large angular spread, \change{in which the changes in $\Hs$ are linear in the current velocity}, and a caustic regime corresponding to strong currents and/or small angular spread. The caustic regime, in which the changes in $\Hs$ are large and concentrated along caustic curves, arises only for parameters values that are outside the range of typical ocean values. We conclude with a summary of our findings and discuss prospects for future work on the spatial variability of $\Hs$ in \S\ref{sec:discussion}.

\section{Formulation}
\label{sec:formulation}

We study the scattering problem sketched  in figure \ref{Fig1}. Deep-water  SGWs, with  small initial directional spreading and a well defined peak frequency (swell)  impinge on a spatially compact coherent flow, such as an axisymmetric vortex or a dipole.

\subsection{Action conservation equation}

\begin{figure}
  \centering
  \includegraphics[trim=12mm 38mm 16mm 8mm, clip,width=0.9\textwidth]{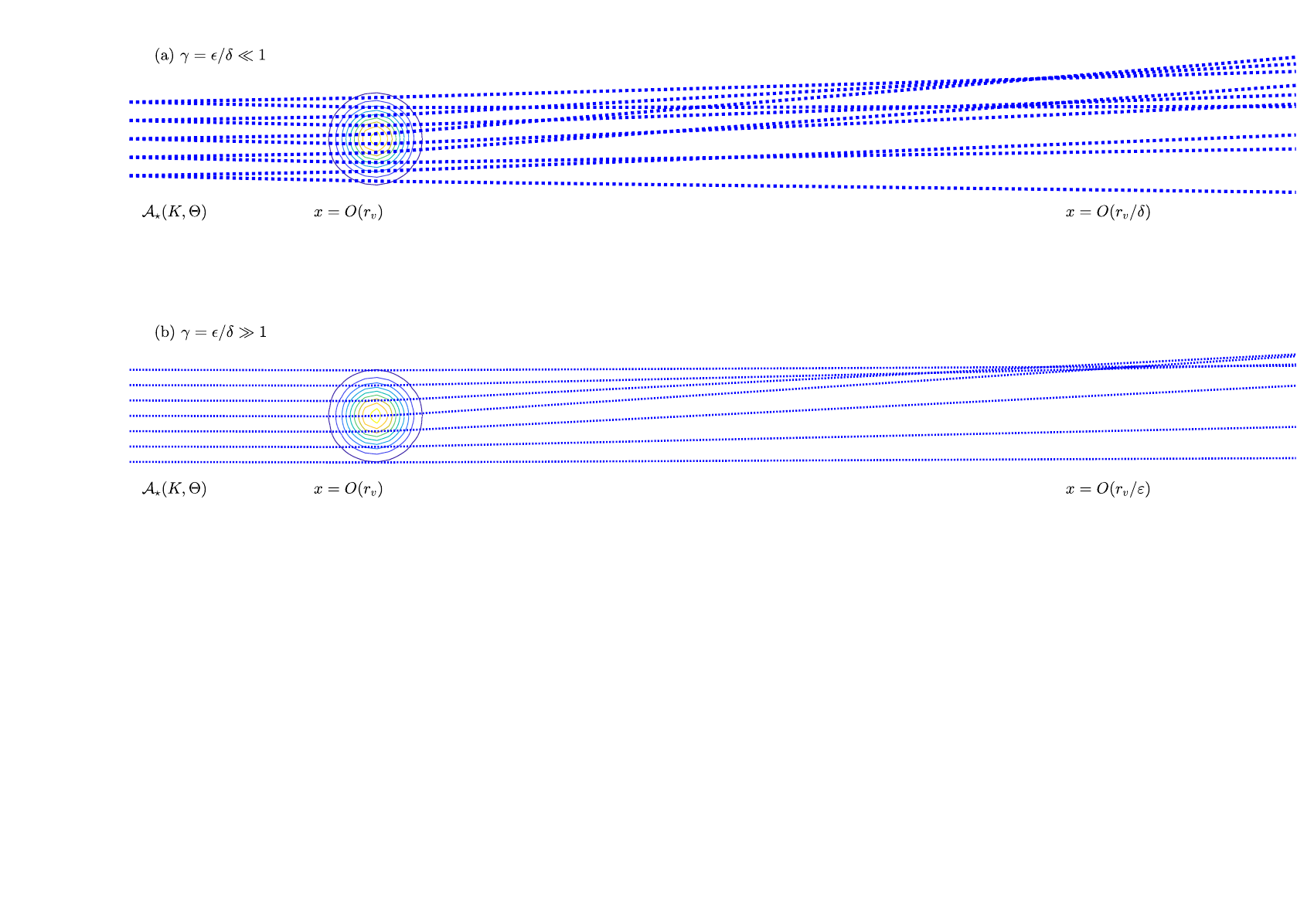}
  \vskip -0.5truein
  \caption{The scattering problem: a localised flow, here shown as an  axisymmetric vortex with radius $\rv$, scatters waves incident from the left ($x\to - \infty$) with  action spectrum $\Acs(K,\Theta)$.  Rays bend significantly only in the scattering region in which there is non-zero vorticity i.e. where $x = O(\rv)$. In this illustration $\rv$ is equivalent to $\ells$. (a) The case $\delta \neq 0$: directional spreading in the incident spectrum $\Acs$  is indicated schematically by two rays emanating from each source point. (b) The case $\delta =0$ (or much less than $\eps$): the incident spectrum $\Acs$ is a plane wave with little or no directional spreading.}
  \label{Fig1}
\end{figure}

 In figure \ref{Fig1} we illustrate the scattering problem by tracing rays through an axisymmetric vortex. We go beyond ray tracing, however,  by using asymptotic methods to obtain approximate analytic  solutions of the conservation equation
\beq
\partial_t \Ac + \bnabla_\bk  \omega \bcdot \bnabla_\bx \Ac - \bnabla_\bx \omega \bcdot \bnabla_\bk \Ac = 0
\label{actioncons}
\eeq
for the wave action density $\Ac(\bx,\bk,t)$  in the four-dimensional position--wavenumber  space \citep{KomenBook,Janssen}. The action conservation equation \eqref{actioncons} relies on the WKB assumption of spatial scale separation between waves and currents.   In \eqref{actioncons} $\omega(\bx,\bk)$ is the absolute frequency of deep-water SGWs
\beq
\omega(\bx,\bk) = \sigma(k) + \bm{k} \bcdot \bU(\bx).
\label{frequency}
\eeq
We consider deep-water waves so that in  \eqref{frequency}  the intrinsic frequency  is $\sigma(k) = \sqrt{g k}$,  with  $k = |\bk|$. The 
current velocity is taken to be horizontal and independent of time and depth,
\beq
\bU(\bx)= U(x,y)\hat\bx + V(x,y) \hat\by .
\eeq

\change{The wave action equation \eqref{actioncons} provides a phase-averaged description of the scattering problem made possible by the scale separation between waves and currents. This places our work in contrast to that of \citet{CosteLund1}, \citet{CosteLund2} and \citet{McIntyre2019} who examined scattering without the simplification afforded by scale separation and discuss phase effects such as the Aharonov--Bohm effect. We also assume fixed currents and do not consider how these might be modified by the presence of waves \cite[see e.g.][]{Humbert2017,McIntyre2019}.}

\subsection{Action spectrum and significant wave height}

Denoting the sea-surface vertical displacement by  $\zeta(\bx,t)$, with root mean square $\zetrms$,  {and following \cite{KomenBook},} we introduce a spectrum $\Fc(\bk,\bx,t)$ such that
\beq
\zetrms^2(\bx,t) =  \int\!\Fc(\bk,\bx,t) \, \dd \bk .
\label{spec3}
\eeq
Later we use  a polar coordinate system $(k,\theta)$ in $\bk$-space so that in \eqref{spec3} $\dd \bk = k \, \dd k \dd \theta$. {The kinetic and potential energy densities for deep-water SGWs are equipartitioned so that  the energy spectrum is $g \Fc$ and the action spectrum, $\Ac(\bx,\bk,t)$ in \eqref{actioncons}, is  $\Ac = g \Fc/\sigma$.} The significant wave height, $4\zetrms$ \citep[][]{KomenBook}, is therefore
\beq
\Hs(\bx,t) =  \left( \frac{16}{g} \int \! \Ac(\bk,\bx,t) \sigma(k) \dd \bk \right)^{1/2}.
\label{stop}
\eeq

The incident swell is characterized by a spatially uniform  spectrum $\Fcs(\bk)$ with constant significant wave height $\Hss$. The subscript $\star$ denotes quantities associated with the incident waves. Swell is  characterized by a narrow spectrum in both wavenumber $k$ (equivalently, frequency $\sigma$) and direction $\theta$. The dominant wavenumber of the incident swell is $\ks$ with frequency $\sigmas = \sqrt{g \ks}$, and the dominant direction is taken without loss of generality as $\theta = 0$. Thus,  as illustrated in figure \ref{Fig1}, the waves arrive from $x = -\infty$ and impinge on an isolated flow feature, centred at $(x,y)=(0,0)$. As an example of  incident spectrum  we use a separable construction described in appendix \ref{modelAppen}.  In the narrow-band limit corresponding to swell,  this spectrum simplifies to the Gaussian
\beq
\Fcs(k,\theta) \approx \zetrmss^2 \, \underbrace{\frac{\ee^{- {(k-\ks)^2}/{2 \delta_k^2}}}{\ks \sqrt{2 \pi \delta_k^2}} }_{F_\star(k)} \times \underbrace{\frac{\ee^{ - \theta^2/2 \delta_\theta^2}}{\sqrt{2 \pi \delta_\theta^2}}}_{D_\star(\theta)}.
\label{LHCS54}
\eeq
The two parameters $\delta_k$ and $\delta_\theta$ capture the wavenumber and directional spreading (see Appendix \ref{modelAppen}). The narrow-band limit assumes that $\delta_k/\ks \ll 1$ and $\delta_\theta \ll 1$.

\section{The scattering problem \label{scatterSec}}

We consider an incident spectrum such as  \eqref{LHCS54}. To make its localisation in  $k$  and $\theta$ explicit we introduce the $O(1)$ independent variables
\beq
K =  \frac{k - \ks}{\del} \quad \textrm{and} \quad  \Th = \frac{\theta}{\del},
\label{KTdefn}
\eeq
where $\del \ll 1$ is a small dimensionless parameter. The incident  action spectrum  has the form
\beq
\Ac(x,y,k,\theta) = \Acs(K,\Theta) \quad \textrm{as} \ \  x \to - \infty,
\label{incident}
\eeq
where the function $\Acs(K,\Th)$ is localised  where both $K$ and $\Th$ are $O(1)$.
The example spectrum \eqref{LHCS54} is  of this form provided that $\delta_k/\ks$ and $\delta_\theta$ are both $O(\delta)$.
This assumption of similarly small spectral widths in $k$ and $\theta$ enforces the relevant distinguished limit for the scattering problem.

We  assume that the currents are weak \citep[e.g.][]{peregrine1976interaction, VillasBoas2020b}. This  means that the typical speed $U$ of the currents is much less than the intrinsic group velocity of the incident swell $\cs$:
\begin{align}
\eps &\defn U/\cs, \label{eps} \\
&\ll 1.
\end{align}
Accordingly we rewrite the frequency \eqref{frequency} as
\beq
\omega(\bx,\bk) = \sigma(k) + \eps \bm{k} \bcdot \bU(\bx). 
\label{freq7}
\eeq
We indulge in a slight abuse of notation here: we develop the approximation  in dimensional variables, hence the dimensionless parameters $\eps$  and $\delta$ in expressions such as \eqref{KTdefn} and \eqref{freq7} should be interpreted as bookkeeping parameters to be set to one at the end. We  examine the distinguished limit
\beq
\delta, \, \eps \to 0 \quad \textrm{with} \quad  \gamma \defn \eps / \delta = O(1) 
\eeq
and  use matched asymptotics   to solve the action conservation equation  \eqref{actioncons}. 
\change{We emphasise that $\gamma = O(1)$ is a formal assumption that enables us to capture the broadest range of relative size of $\eps$ and $\delta$, including $\eps \ll \delta$ and $\delta \ll 1$ (see \S\ref{sec:limits}).}

\subsection{The scattering region: $x = O(\ells)$}

The spatially compact flow  has a typical horizontal length scale which we denote by $\ells$.
We refer to the region where  $x=O(\ells)$ as  the `scattering region'. The solution in this region has the form
\beq
\Ac(K,\Th,x,y)
\label{ansatz3}
\eeq
and must limit to $\Acs(K,\Th)$ in \eqref{incident} as $x \to -\infty$. 

With $\Ac$ in \eqref{ansatz3}  the transport term in \eqref{actioncons} is approximated as 
\begin{align}
\bnabla_\bk  \omega \bcdot \bnabla_\bx \Ac &= \cs \left( \cos(\delta \Th) \Ac_x + \sin (\delta \Th) \Ac_y \right) 
+ \eps \bU\bcdot \bnabla_{\bx} \Ac \nonumber \\
&= \cs \Ac_x + O(\delta,\eps).
\label{bareg3}
\end{align}
In particular, transport by the current, $\eps \bU\bcdot \bnabla_{\bx} \Ac$ is negligible compared with transport by the intrinsic group velocity $\cs$.
With the approximations 
\begin{align}
\bnabla_{\bk} \Ac &= \delta^{-1}  \left( \partial_K \Ac \, \hat\bx + k_*^{-1} \partial_\Theta \Ac \, \hat \by \right) +O(1), \label{pkA}\\
 \bnabla_\bx  \omega &= \eps \ks (U_x \hat\bx + U_y  \hat{\bm{y}}) +O(\eps \delta),
\end{align}
the refraction term in \eqref{actioncons} simplifies  to
\beq
\bnabla_\bx \omega \bcdot \bnabla_\bk \Ac =  \gamma \left( \ks U_x \partial_K \Ac + U_y \partial_\Theta \Ac \right) + O(\eps).
\eeq 
Thus in the scattering  region the  leading-order approximation to  \eqref{actioncons} is
\beq
\cs \partial_x \Ac - \gamma \left( \ks U_x \partial_K \Ac + U_y \partial_\Theta \Ac \right) = 0,
\label{xO1}
\eeq
\change{One might solve \eqref{xO1} using its characteristics -- the ray equations -- or  by inspection.
By either method the solution to \eqref{xO1} that matches the incident action spectrum  \eqref{incident} as $x \to - \infty$ is found to be}
\beq
\Ac(x,y,K,\Theta) = \Acs \left(K + \frac{\gamma  \ks}{ \cs} U(x,y)\, , \Theta+\frac{\gamma}{\cs}  \int_{-\infty}^x \!\!\!  U_y(x',y) \, \dd x' \right).
\label{inner}
\eeq
It is insightful to introduce the vorticity $Z \defn V_x - U_y$ and write  \eqref{inner} as
\beq
\Ac(x,y,K,\Theta) = \Acs \left(K + \frac{\gamma  \ks}{ \cs} U(x,y) \, , \Theta+ \frac{\gamma}{\cs}  V(x,y) -  \frac{\gamma}{\cs}\int_{-\infty}^x \!\!\!Z(x',y)\, \dd x' \right).
\label{inner3-1}
\eeq
For reference, we rewrite this expression in terms of the original independent variables, setting the bookkeeping parameters $\eps$, $\delta$, and hence $\gamma$ to $1$ to obtain
\beq
\Ac(x,y,k,\theta) = \Acs \left(k + \frac{\ks}{ \cs} U(x,y) \, , \theta+ \frac{1}{\cs}  V(x,y) -  \frac{1}{\cs}\int_{-\infty}^x \!\!\!Z(x',y)\, \dd x' \right).
\label{inner3}
\eeq

\subsection{The intermediate region: $O(\ells)\ll x \ll O(\ells/\delta)$}

The outer limit of the  inner solution  \eqref{inner3-1} follows from taking $x \to \infty$:  
\beq
\Ac(x,y,K,\Theta) \to  \Acs\left(K, \, \Theta - \gamma \Delta(y) \right),
\label{match1}
\eeq
where we have introduced the dimensionless  `deflection'
\beq
\Delta(y) \defn \frac{1}{\cs}\int_{-\infty}^{\infty} \!\!\!Z(x',y)\, \dd x' .
\label{inner7}
\eeq
According to  \eqref{match1} the effect of the flow on the dependence of $\Ac$ on $K$ is reversible:  after passage through the scattering region  this dependence reverts to the incident form. 
\change{In contrast, there is a net change in $\Theta$, quantified by the deflection $\Delta(y)$. This can be related to classical scattering of particles by viewing $y$ as the impact parameter of a wavepacket. The scattering cross section, defined as $\dd y / \dd \theta_\infty$ where $\theta_\infty$ is the angle of propagation of the wavepacket as $x \to \infty$, is then $-1/(\eps \Delta'(y))$.}

To physically interpret \eqref{match1} and $\Delta(y)$,  recall that if  $\eps$ is small then
\begin{align}
\text{ray curvature} &\approx \frac{\text{vorticity}}{\text{group velocity}}\, , \label{curvature3}\\
 &\approx \frac{Z(x,y)}{\cs}.
 \label{curvature33}
\end{align}
The approximation in \eqref{curvature3} requires only $\eps \ll 1$ \citep[e.g.][]{Kenyon1971,LLfluid,Dysthe2001,Gallet2014}. Passing from \eqref{curvature3} to \eqref{curvature33} requires the further approximation that $k$ is close to $\ks$ so that the group velocity in the denominator of \eqref{curvature3} can be approximated by the constant $\cs$. On the left of \eqref{curvature3} ray curvature is $\dd\theta/\dd \ell$, where $\ell$ is arc-length along a ray. But within  the compact scattering region we   approximate $\ell$ with $x$.  Thus the deflection $\Delta(y)$ in \eqref{inner7}  is the integrated ray curvature, accumulated as rays pass through the scattering region in which $x=O(\ells)$ and vorticity $Z(x,y)$ is non-zero.

From \eqref{inner7} and \eqref{curvature3} we conclude that  the scattering region is best characterized as the region with $O(1)$ vorticity, e.g.\ the vortex core in figure \ref{Fig1} (hence $\ells = \rv$ with $\rv$ a typical vortex radius).  The region with palpably non-zero velocity  is much larger. In figure \ref{Fig1} the rays are straight  where $x=O(\rv/\eps)$, despite the slow ($\propto r^{-1}$) decay of the  azimuthal vortex velocity.

\subsection{The far field: $x = O(\ells/\delta)$}
 Far from the scattering region, where $x \gg \ells$, we introduce the slow coordinate  $X\defn\delta x$. In the far-field  the currents  and hence  the refraction term $\bnabla_\bx \omega \bcdot \bnabla_\bk \Ac$ in \eqref{actioncons} are negligible. 
The steady action conservation equation collapses to 
\beq
\bnabla_{\bk} \sigma \bcdot \bnabla_{\bx} \Ac = \cs \left( \delta \cos(\delta \Th) \Ac_X + \sin(\delta \Th) \Ac_y \right) =0,
\eeq
i.e. propagation along straight rays. Retaining only the leading-order term gives
\beq
 \partial_X \Ac +  \Theta \partial_y \Ac = 0,
 \label{XO1}
\eeq
By inspection the  solution of \eqref{XO1} that matches the intermediate solution  \eqref{match1} is 
\beq
\Ac(X,y,K,\Theta) = \Acs \left(K, \Theta - \gamma  \Delta\left(y -  X \Theta \right)\right).
\label{AcXO1-1}
\eeq
This formula, which converts the incident spectrum into the far-field spectrum, is a key  result of the paper. In terms of the original independent variables and with the bookeeping parameters set to $1$ it takes the convenient form
\beq
\Ac(x,y,k,\theta) = \Acs \left(k, \theta -   \Delta\left(y -  x \theta \right)\right).
\label{AcXO1}
\eeq

\subsection{Significant wave height}

Significant wave height $\Hs$ is the most commonly reported statistic of wave amplitudes, being routinely observed by satellite altimeters and wave buoys. We obtain an approximation for $\Hs$ by performing the $k$ and $\theta$ integrals  in  \eqref{stop}  using  the approximations \eqref{inner3} and \eqref{AcXO1} for $\Ac(\bx,\bk)$. 

The scattering region is simple. We can approximate $\sigma$ and $\dd \bk$ in \eqref{stop} by $\sigmas= \sigma(\ks)$ and $\ks \, \dd k \dd \theta$ to find
\begin{align}
\Hs(\bx,t) &\sim  \left( \frac{16 \sigmas \ks}{g} \iint \!\Ac(\bk,\bx,t) \dd k \dd \theta\right)^{1/2} \label{HsSc} \\
&\sim \Hss 
\end{align}
 The second equality holds because, according to \eqref{inner3}, $\Ac(\bx,\bk)$ is obtained from $\Acs(\bx,\bk)$  by an $\bx$-dependent shift of the $k$ and $\tha$ that does not affect the integral. Thus $\Hs$ in the scattering region  is unchanged from the incident value $\Hss$. This conclusion also follows  directly from steady-state wave action conservation under the assumptions $\eps, \, \delta \ll 1$: multiplying \eqref{xO1} by $\sigmas \ks$ and integrating over $k$ and $\theta$ we find
\beq
\cs \p_x  \underbrace{  \left( \sigmas  \ks \iint \! \Ac(\bx,\bk)  \, \dd k \dd \theta \right) }_{\approx g \Hs^2(\bx) /16}  =0. \label{innerHs}
\eeq
Hence $\Hs(\bx) = \Hss$  throughout the scattering region.

In the far field, $\Hs$ is obtained  by substituting \eqref{AcXO1} into \eqref{stop}. The result is 
\beq
\Hs(\bx) = 4 \sqrt{\frac{\ks \sigmas}{g}  \int \!\! \dd \theta \int \!\!\dd k \, \Acs(k,\theta -  \Delta(y - x \theta)) }.
\label{farStop}
\eeq
The $k$-integral can be  evaluated in terms of the incident directional spectrum which, in the general case of a non-separable spectrum, is defined as
\beq
D_\star(\tha) \defn \frac{1}{\zetrmss^2} \int\!\Fcs(\bk) \,k \, \dd k. \label{dirspec}
\eeq
We summarize the results above with:
\beq
\Hs(\bx) = \Hss \begin{cases} 1  & \text{in the scattering region,}\\
\sqrt{\int \Ds\left(\theta - \Delta (y-x \theta)\right)\dd \theta} & \text{in the far field.} \label{HsSumm}
\end{cases} 
\eeq

\section{Applications to simple flows \label{simpleFlows}}

\subsection{Gaussian vortex} \label{sec:gaussian}

As an application, we consider  scattering by  an axisymmetric  Gaussian vortex with  circulation $\kappa$, vorticity
\beq
Z(x,y) = \frac{\kappa\, \ee^{-r^2/2 \rv^2}}{2 \pi r_v^2} ,
\label{gaussian}
\eeq
and velocity
\beq
\left(U(x,y),V(x,y)\right) = \frac{\kappa}{2 \pi } \frac{1 -  \ee^{-r^2/2 \rv^2} }{r^2} \left(-y,x\right),
\eeq
where $r^2 = x^2+y^2$. The vortex radius $\rv$ can be taken as the scattering length scale $\ells$. The maximum azimuthal velocity is $U_m = 0.072 \, \kappa/\rv$ at radius $1.585 \, \rv$. The deflection \eqref{inner7} resulting from this Gaussian vortex is
\beq
\Delta(y)=  \frac{\kappa \, \ee^{-y^2/2\rv^2} }{\sqrt{2 \pi}\,  \rv \cs} .
\label{Del77}
\eeq

The asymptotic solution in the scattering region is obtained from \eqref{inner3} as
\begin{align}
\Ac(x,y,k,\tha) &= \Acs  \Big( k +   k_* c_*^{-1} U(x,y),  \nonumber \\ 
 & \qquad  \qquad \tha + c_*^{-1} V(x,y) -  \tfrac{1}{2} \left( \textrm{erf}\big({x}/{\sqrt{2} r_v} \big) + 1\right) \Delta(y)  \Big),
\label{AcGVSc}
\end{align}
where $\textrm{erf}$ is the error function. 
Eq.\ \eqref{AcGVSc} can be combined with the far-field approximation \eqref{AcXO1}  into a single, uniformly valid approximation,
\begin{align}
\Ac(x,y,k,\tha) &= \Acs  \Big( k +   k_* c_*^{-1} U(x,y),  \nonumber \\ 
 & \qquad  \qquad \tha + c_*^{-1} V(x,y) -  \tfrac{1}{2} \left( \textrm{erf}\big({x}/{\sqrt{2} r_v} \big) + 1\right) \Delta(y - x \tha)  \Big). \label{uniform}
\end{align}
The significant wave height is approximated by \eqref{HsSumm} which can be written as the uniform expression
\beq
\Hs(x,y) = \Hss \sqrt{\int \Ds\left(\theta - \Delta (y-x^+ \theta)\right)\dd \theta},
\label{uniformHs}
\eeq
where $x^+$ is equal to $x$ for $x>0$ and to $0$ for $x < 0$ and  \eqref{Del77} is used for $\Delta$.

We now compare the matched asymptotic (MA hereafter) predictions \eqref{uniform}--\eqref{uniformHs} with  numerical solutions of the wave action equation \eqref{actioncons} obtained with the Wave Height, Water Depth, and Current Hindcasting  third generation wave model (WAVEWATCH III, hereafter WW3). The incident spectrum used for WW3 is described in Appendix \ref{modelAppen}. The directional function for this spectrum is the \citet{LHCS1963} model
\beq
\Ds(\theta) \propto \cos^{2s} \frac{\theta}{2}.
\label{LHdir}
\eeq
The parameter $s>0$ controls the directional spreading: for $s \gg 1$, \eqref{LHdir} reduces to the Gaussian in \eqref{LHCS54} with directional spreading $\delta_\theta = \sqrt{2/s}$. The configuration of WW3 and spectrum parameters are detailed in Appendix \ref{app:WW3}. The most important parameter is the peak frequency of the incident spectrum, taken fixed for all simulations as $\sigmas = 0.61$ rad s$^{-1}$. This corresponds to a period of 10.3 s, wavelength of 166 m and group speed $\cs = 8$ m s$^{-1}$.  Because  the problem is linear in the action density,  the values of $\zetrmss$ or equivalently $\Hss$ are less important. For definiteness we set $\Hss = 1$ m.  

\begin{figure}
  \centering
  \includegraphics[width=1.0\textwidth]{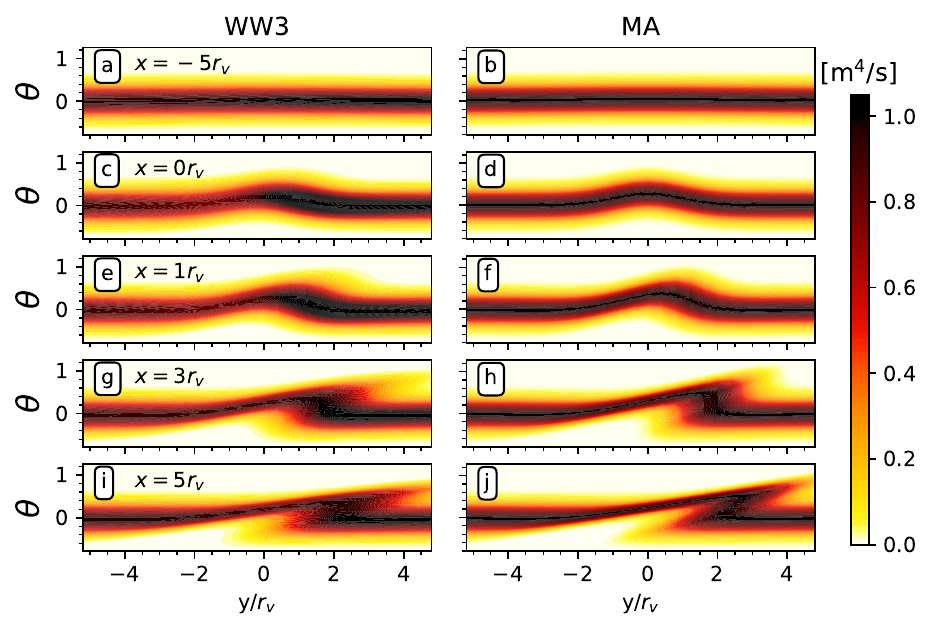}
  \caption {Wavenumber-integrated action density $\int \mathcal{A}(x,y,k,\theta) \, \dd k$ as a function of $y$ and $\theta$ 
  at $x=-5 \, \rv$,  $0$, $\rv$, $3\, \rv$ and $5 \, \rv$ from WW3 (left) and MA  (Eq.\ \eqref{uniform}, right) for swell impinging on a Gaussian vortex with $U_m=0.8$ m s$^{-1}$. The directional spreading of the incident spectrum is $s=40$. }
  \label{A_comparison}
\end{figure}

Figure \ref{A_comparison} compares the wavenumber-integrated  wave action $\int \mathcal{A}(x,y,k,\theta) \, \dd k$ obtained from \eqref{uniform} and WW3 for a Gaussian vortex with maximum velocity $U_m = 0.8$ m s$^{-1}$ and directional spreading parameter $s = 40$. Figure  \ref{A_comparison} shows a good agreement, especially in the far-field region ($x \ge 3 \rv$). The most noticeable difference between MA and WW3 is in  panels c and d, which show a section through the  middle of the vortex.  The MA action spectrum in panel d is obtained via a $y$-dependent shift in $\Acs(k,\tha)$; there is no change in the intensity of $\Ac$ associated with this shift. In panel c, on the other hand,   the intensity of the WW3 action spectrum varies with $y/\rv$. We attribute this difference to  asymptotically small effects such as  the contribution $\bU \bcdot \bnabla_{\bx} \Ac$ to wave-action transport.

\begin{table}
  \begin{center}
\def~{\hphantom{0}}
  \begin{tabular}{ccccc}
      $s$ & $U_m$ (m s$^{-1}$)	 & $\delta=\sqrt{2/s}$          & $\eps=U_m/\cs$ & $\gamma = \eps/\delta$\\ \hline
       10 & 0.4  		          & $0.447\ (25.6^{\circ})$     & 0.05      	     & 0.112\\
       40 & 0.4   		          & $0.224\ (12.8^{\circ})$     & 0.05      	     & 0.224\\
       10 & 0.8 		          & $0.447\ (25.6^{\circ})$     & 0.1   	              & 0.224\\
       40 & 0.8 		          & $0.224\ (12.8^{\circ})$     & 0.1  		     & 0.447\\
  \end{tabular}
  \caption{Parameters corresponding to each configuration in section \ref{sec:gaussian}, arranged in the order of the rows in figure \ref{GV_hs}. In all cases the group speed is $c_\star = 8$ m s$^{-1}$, corresponding to a 166 m wavelength and $10.3$ s period. $U_m$ is the maximum vortex velocity and the vortex radius is $\rv = 25 $ km.}
  \label{tb:parameters}
  \end{center}
\end{table}

In the remainder  of this section, we assess the dependence of significant wave height  $\Hs$ on the directional spreading parameter $s$ and flow strength $U_m$.  We consider the four different combinations of $s$ and $U_m$ given in Table \ref{tb:parameters}. The corresponding values of the dimensionless parameters,  taken as 
\beq
\delta = \delta_\theta  = \sqrt{2/s} \quad \text{and} \qquad \eps = U_m/\cs,
\eeq
are also in the table. 

Observations of the directional spreading for swell typically range between $10^\circ-20^\circ$ \citep{ewans2002directional}, which correspond to a range of $s$ between $16$ and $66$. In our experiments, setting $s=10$ and $s=40$ leads to a directional spreading of $24^\circ$ and $12^\circ$ respectively, which correspond to very broad and very narrow swells.

\begin{figure}
  \centering
  \includegraphics[width=1.0\textwidth]{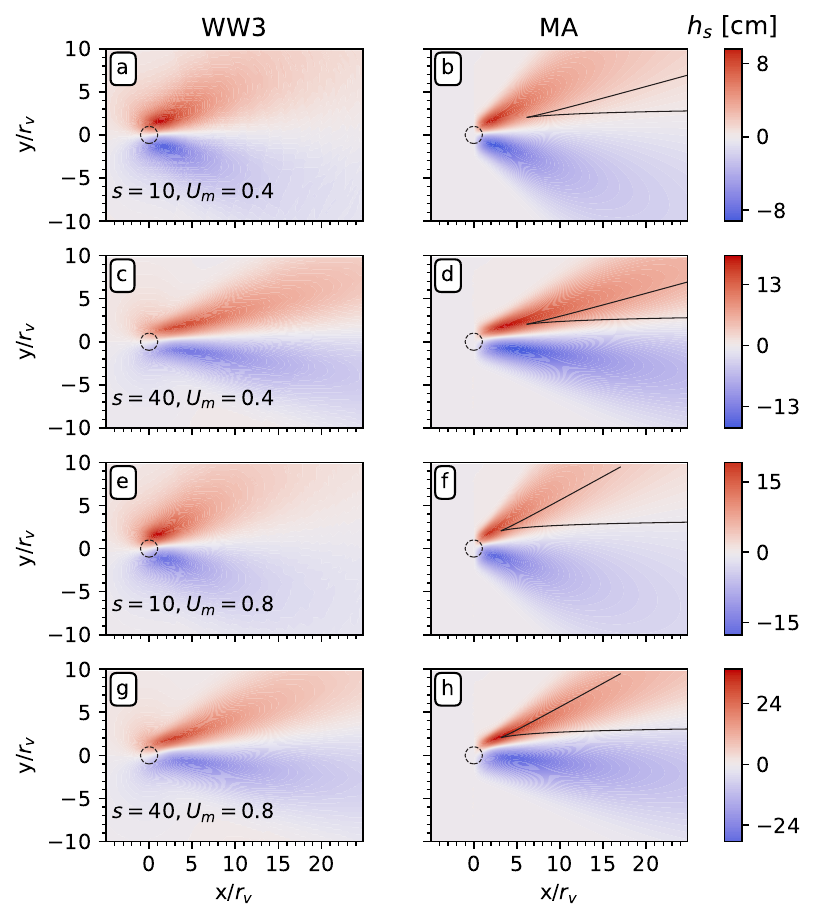}
  \caption {Significant wave height anomaly $\hs(x,y)$ from WW3 (left column) and MA (right column) for swell impinging on a Gaussian vortex. Each row corresponds to the indicated values of the directional spreading parameter $s$ of the incident wave spectrum and of the maximum velocity  $U_m$ (in  m s$^{-1}$). The corresponding non-dimensional parameters are given in Table \ref{tb:parameters}. The dashed circles has radius $\rv$ around vortex center. The solid lines on the right panels indicate the caustics computed from \eqref{causyx}.  The colourbars differ between rows but are the same within each row. White corresponds to $\hs=0$ in all panels.  The \change{customizable} notebook that generates panel (h) \change{by default} can be accessed at \url{https://www.cambridge.org/S0022112023006869/JFM-Notebooks/files/Figure-3.ipynb}. }
  \label{GV_hs}
\end{figure}

\begin{figure}
  \centering
  \includegraphics[width=1.0\textwidth]{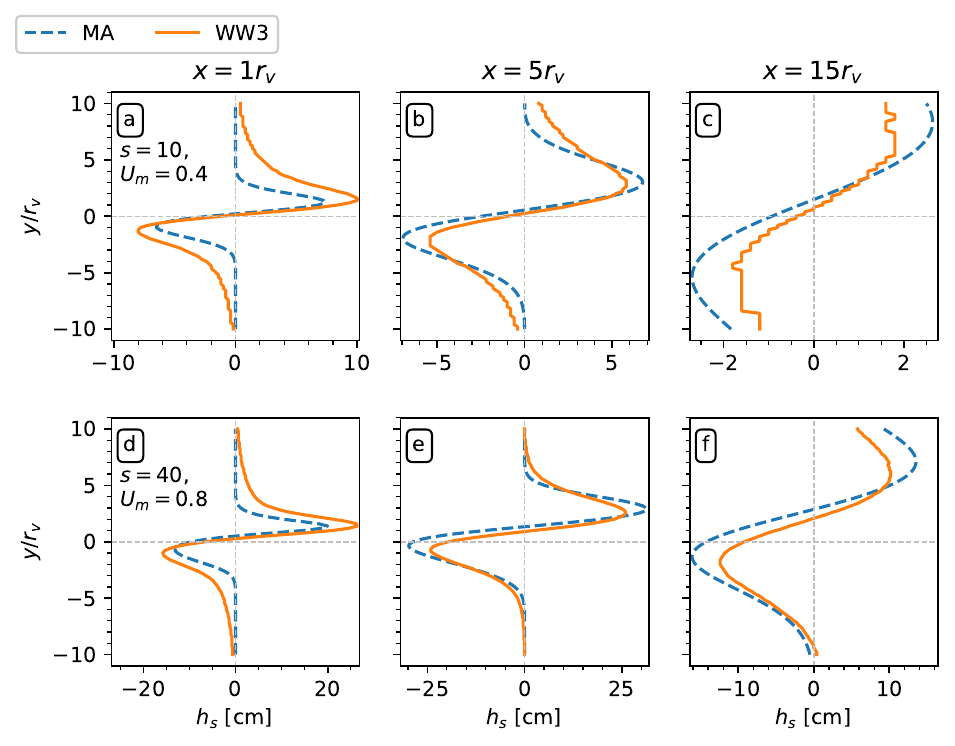}
  \caption {Significant wave height anomaly $h_s$ as a function of $y$ for $x=r_v, 5 \, r_v, 15 \, r_v$ (left, centre and right) from WW3 (solid lines) and MA  (Eq.\ \eqref{uniform}, dashed lines) in the set up of figure \ref{GV_hs}. Results are shown for two sets of parameters $s$ and $U_m$ as indicated in the leftmost panels. The range of $h_s$ differs between panels. }
  \label{GV_crosssec}
\end{figure}

Figures \ref{GV_hs} and \ref{GV_crosssec} show the significant wave height anomaly 
\beq
\hs(\bx)\defn \Hs(\bx)-\Hss
\label{hsdef}
\eeq
for each combination of $s$ and $U_m$.  Because of our choice of $\Hss = 1$m, $\hs$  in cm can  be interpreted as the fractional change in significant wave height expressed as a percentage.
A control run of WW3 in the absence of currents shows that $\hs$ is not exactly zero but decreases slowly with $x$. This is caused by the finite $y$-extent of the computational domain which leads to a wave forcing with compact support. To mitigate this numerical artefact, we compute the WW3 significant wave height anomaly as $\hs(\bx) = \Hs(\bx)-\Hs^{\mathrm{ctrl}}(\bx)$, where $\Hs^{\mathrm{ctrl}}(\bx)$ is the significant wave height of the current-free control run. See Appendix \ref{app:WW3} for details.

Figures  \ref{GV_hs} and \ref{GV_crosssec}  show that $\hs$ has a wedge-like pattern in the  wake of the vortex resulting from wave focussing and defocussing, with $\hs > 0$ mainly for $y > 0$ and $\hs < 0$  for $y < 0$. The pattern is not anti-symmetric about $y=0$, and positive anomalies are larger than negative anomalies. These characteristics, which indicate a nonlinear response, are increasingly marked as $s$ and $U_m$ increase. Specifically, the parameter
\beq
\gamma = \frac{\eps}{\delta}  = \frac{U_m}{\cs} \sqrt{\frac{s}{2}} \label{gammaest}
\eeq
controls the degree of nonlinearity and hence of asymmetry. We discuss the two limiting regimes $\gamma \ll 1$ and $\gamma \gg 1$ in \S\ref{sec:limits}.

There is good overall agreement between WW3 and MA, even though, in the case $s=10$, the parameter $\delta = 0.447$ is only marginally small. The pattern is more diffuse for WW3 than for MA, with a less sharply defined wedge and a non-zero $\hs$ over a larger proportion of the domain. We attribute the differences to the finiteness of $\delta$ (they are more marked for $s=10$, $\delta = 0.447$ than for $s=40$, $\delta=0.224$), and to the limited spectral resolution of WW3 (simulations with degraded angular resolution lead to an even more diffuse $\hs$).
The most conspicuous differences between WW3 and MA appear in the scattering region, where the non-zero $\hs$ obtained with WW3 appears to contradict the MA prediction that $\hs = 0$. The non-zero $\hs$ results from $O(\eps,\, \delta)$ terms neglected by MA. Relaxing some of the approximations
leading to \eqref{HsSc}  gives a heuristic correction to MA that captures the bulk of the difference with WW3 in the scattering region. 
We explain this in Appendix \ref{app:mismatch}.

As further demonstration of the MA approach, we provide a Jupyter notebook accessible at \url{https://www.cambridge.org/S0022112023006869/JFM-Notebooks/files/Figure-3.ipynb}, where users can \change{customize} the form of the current and the incoming wave spectrum to experiment with the resulting  $\int \mathcal{A}(x,y,k,\theta) \, \dd k$  and $\hs$.

\subsection{Vortex dipole}

\begin{figure}
  \centering
  \includegraphics[width=1.0\textwidth]{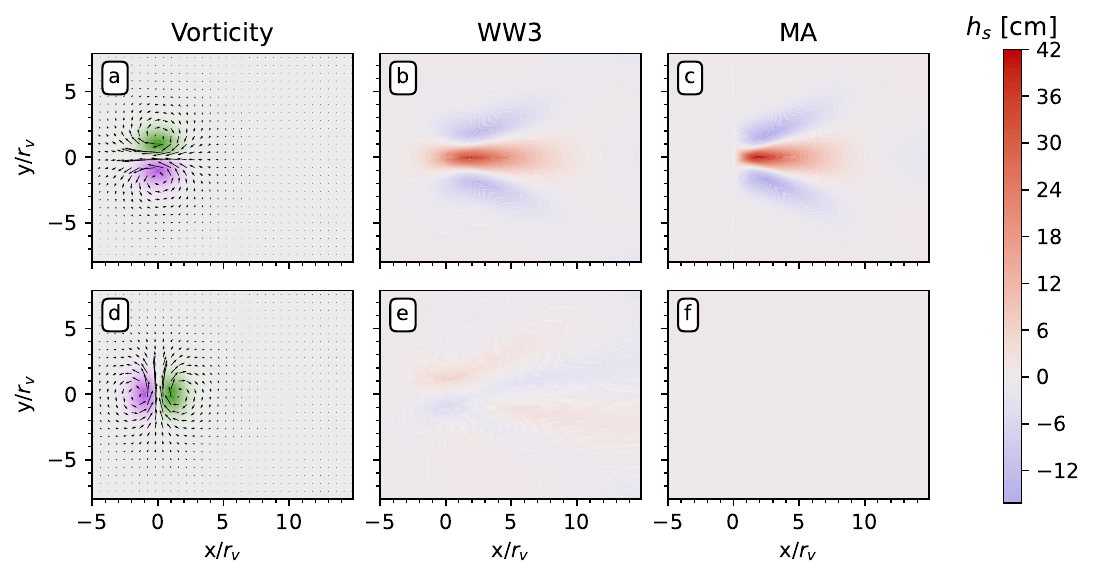}
  \caption{Swell impinging on vortex dipoles with axes perpendicular (top) and parallel (bottom) to the dominant direction of wave propagation ($x$-axis). The vorticity (colour) and velocity (vectors) are shown (left) together with the significant wave height anomaly $\hs$ from WW3 (middle) and MA (right). The directional spreading parameter $s = 40$ and the maximum flow velocity is $0.8$~m~s$^{-1}$.}
  \label{fig:dipole}
\end{figure}

A striking feature of the far-field spectrum and hence of $\Hs$  is that, according to MA, they depend on the flow only through the deflection $\Delta(y)$ in \eqref{inner7}, proportional to the integral of the vorticity along the direction of dominant wave propagation (the $x$-direction in our set up). This implies that if the integrated vorticity vanishes because of cancellations between positive and negative contributions, the differences between far-field and incident fields are asymptotically small. This can be interpreted as a form of `vortex cloaking', whereby an observer positioned well downstream of a flow feature is unable to detect its presence through changes in wave statistics. We demonstrate this phenomenon by examining the scattering of swell by vortex dipoles. 

We consider two cases, corresponding to dipoles whose axes (the vector joining the centres of positive and negative vorticity) are, respectively, perpendicular and parallel to the direction of wave propagation.  
The corresponding vorticity fields are chosen, up to a constant multiple, as the derivative of the Gaussian profile \eqref{gaussian} with respect to $y$ or $x$. Figure \ref{fig:dipole} shows the significant wave height anomaly obtained for the incident spectrum of \S\ref{sec:gaussian} with $s=40$ and dipoles with maximum velocity $U_m=0.8$ m s$^{-1}$. 

When the dipole axis is in the  $y$-direction (top row) the deflection $\Delta(y)$ does not vanish identically. As a result, $\Hs$ is affected by the flow, strongly for our choice of parameters. This applies to both the MA and WW3 predictions which match closely in the far field. When the dipole axis is in the $x$-direction (bottom row), $\Delta(y) = 0$. The MA  prediction is then that $\Hs = \Hss$, i.e.\ $\hs = 0$, everywhere. The WW3 simulation is consistent with this, with only a weak signal in $\hs$.

In general, for a dipole with axis making an angle $\alpha$ with the direction of wave propagation, the deflection $\Delta(y)$ is proportional to $\sin \alpha$ and the cloaking effect is partial unless $\alpha = 0$.

\section{Limiting cases} \label{sec:limits}

In this section, we return to the far-field asymptotics \eqref{AcXO1-1} for $\Ac$ in terms of the scaled dependent variables in order to examine two limiting regimes characterized by extreme values of $\gamma = \eps / \delta$. The regime $\gamma \ll 1$ corresponds to a weak flow and/or relatively broad spectrum, leading to a linear dependence of $\hs$ on the \change{currents}. The opposite regime $\gamma \gg 1$ corresponds to strong flow and/or highly directional spectrum. The wave response is then highly nonlinear \change{in the currents} and, as we show below, controlled by the caustics that exist for pure-plane incident waves ($\gamma = \infty$). \citet{Heller2008}'s `freak index', given by $\eps^{2/3}/\delta$, is the analogue of $\gamma$ for spatially extended, random currents.  

\subsection{Linear regime: $\gamma \ll 1$}\label{sec_lin}

For $\gamma \ll 1$, we can expand \eqref{AcXO1-1} in Taylor series to obtain
\beq\label{Ataylor}
\Ac(X,y,K,\Theta) = \Acs(K, \Theta)  - \gamma  \Delta(y -  X \Theta)\,  \partial_\Theta \Acs(K, \Theta) + O(\gamma^2).
\eeq
This indicates that the flow induces the small correction $- \gamma  \Delta(y -  X \Theta) \partial_\Theta \Acs(K, \Theta)$ to the action of the incident wave. We  deduce an approximation for $\Hs$ by integrating \eqref{Ataylor} with respect to $K$ and $\Theta$ to obtain $\Hs^2$ followed by a Taylor expansion of a square root. Alternatively, we can carry out a Taylor expansion of the far-field approximation \eqref{HsSumm} of $\Hs$, treating $\Delta(y)$ as small. The result is best expressed in terms of the anomaly $\hs$, found to be 
\beq
\hs(x,y) = - \frac{H_{s\star}}{2}  \int  D_s'(\theta) \, \Delta(y-x \theta) \, \dd \theta \label{hstaylor}
\eeq
after reverting to the unscaled variables and setting $\gamma = 1$. This simple expression is readily evaluated once the flow, hence $\Delta(y)$, and directional spectrum $\Ds(\theta)$ are specified. For the Gaussian vortex of \S\ref{sec:gaussian}
and the directional spectrum in \eqref{LHCS54}, the integration can be carried out explicitly, yielding
\beq\label{hslinear}
\hs(x,y)= \frac{\Hss \kappa}{\cs \sqrt{\pi}} \frac{x^+ y\,  \ee^{-y^2/\LL(2r_v^2+4x^2/s\RR)}}{(2 r_v^2+4x^2/s)^{3/2}}.
\eeq
This formula makes it plain that $\hs$ depends on space through $(x/\sqrt{s}, y)$, is antisymmetric about the $x$ axis, and is maximised along the curves $y=\pm\sqrt{r_v^2+2x^2/s}$. Decay as $|\bx| \to \infty$ is slowest along these curves and proportional to $x^{-1}$.

\begin{figure}
  \centering
  \includegraphics[width=1.0\textwidth]{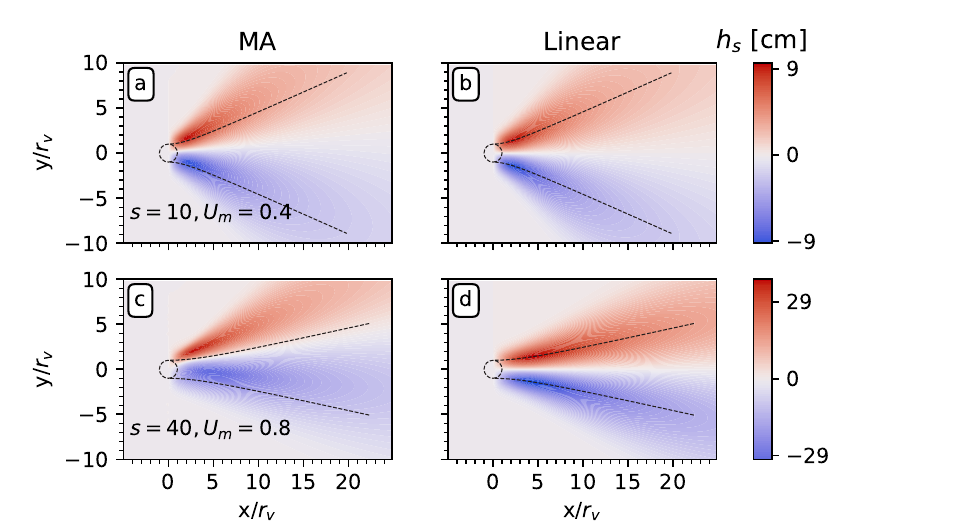}
  \caption {Significant wave height anomaly  $\hs(x,y)$ for swell impinging on a Gaussian vortex: comparison between the predictions of
  MA (left) and and its $\gamma \to 0$ limit (\eqref{hslinear}, right column). The set up is as in figure \ref{GV_hs} with parameters $s$ and $U_m$ (in m s$^{-1}$) as indicated. Dashed lines indicates the curves $y=\pm\sqrt{r_v^2+2x^2/s}$ where $\hs$ reach maximum amplitudes according to \eqref{hslinear}.}
  \label{fig_hsMAlinear}
\end{figure}

We illustrate \eqref{hslinear} and assess its range of validity by comparing it with MA for two sets of parameters in figure \ref{fig_hsMAlinear}. The match is very good for $s=10$ and $U_m = 0.4$ m s$^{-1}$ (top row), corresponding to
$\gamma = 0.112$. It is less good for $s=40$ and $U_m = 0.8$ m s$^{-1}$, unsurprisingly since $\gamma = 0.447$ is not particularly small and the MA prediction is obviously far from linear, with a pronounced asymmetry. 
The curves $y=\pm\sqrt{r_v^2+2x^2/s}$ shown in the figure are useful indicators of the structure of $\hs$ for small enough $\gamma$.

\subsection{Caustic regime: $\gamma \gg 1$} 

\begin{figure}  
 \includegraphics[width=\textwidth]{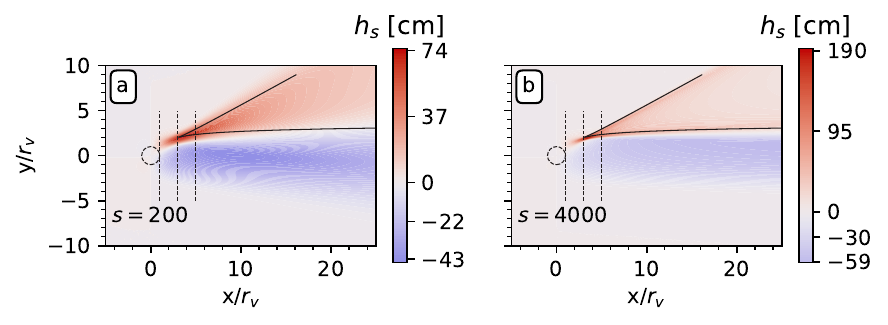}
 \caption{Caustics for swell impinging on a Gaussian vortex: the caustics \eqref{causyx} (solid lines) are superimposed to the MA prediction of $\hs$ for $U_m=0.8$ m s$^{-1}$ and the indicated values of $s$. The dashed vertical lines correspond to the values of $x = r_v$, $3\, r_v$ and $5 \, r_v$ used in figure \ref{fig:caustics2}.}
 \label{fig:caustics1}
\end{figure}

The limit $\gamma \to \infty$ corresponds to an incident wave field that is almost a plane wave.
It is natural to rescale variables according to $\Theta \mapsto \gamma \Theta$ and $X \mapsto \gamma^{-1} X$ so that \eqref{AcXO1-1} becomes
\beq
\Ac(X,y,K,\Theta) = \Acs\left(K, \gamma \Sc(X,y,\Theta) \right),
\label{Accaustic}
\eeq
where
\begin{align}
\Sc(X,y,\Theta) \defn \Theta -  \Delta(y -  X \Theta).
\label{surface}
\end{align}
In $(X,y,\Theta)$-space, the $K$-integrated action is  concentrated  in a thin $O(\gamma^{-1})$ layer around the surface $\Sc(X,y,\Th)=0$.
 Quantities such as $\Hs$ obtained by further integrating the action with respect to $\Theta$ can be obtained by approximating the dependence of right-hand side of \eqref{Accaustic} on $\Sc$ by $\delta( \Sc)$. This fails, however, when $(X,y,\Theta)$ satisfy both 
\beq\label{cloc}
\Sc(X,y,\Theta) = 0, \qquad \text{and} \qquad \partial_\Theta \Sc(X,y,\Theta)=  1 + X  \Delta'( y -  X \Theta) = 0.
\eeq
The corresponding curves in the $(X,y)$ plane are caustics near which $\int \Ac(X,y,K,\Theta) \, \dd K  \dd \Theta$ is an order $\gamma^{1/2}$ larger than elsewhere; correspondingly, $\Hs = O(\gamma^{1/4})$.  In figure \ref{fig:caustics1} the two caustics  meet at a cusp point from  opposite sides of a common tangent.  The cusp point is located by the  condition $\partial^2_{\Theta}  \Sc = 0$  and the integrated action at the cusp point is $O(\gamma^{2/3})$ so that $\Hs = O(\gamma^{1/3})$.  We have numerically verified these  $\gamma$-scalings at the caustics and at the cusp point   by varying $s$ in the MA solutions.

For the Gaussian vortex \eqref{gaussian}, the system \eqref{cloc} can be solved to obtain an explicit equation for the caustics. This equation is derived in Appendix \ref{app:caustics} and given by \eqref{causyx}. It describes two curves $y(x)$ emanating from the cusp point at $x = x_c$ given by  \eqref{xc}. The caustics (which depend on $U_m$ but not on $s$) are indicated on the right panels of figure 
\ref{GV_hs}. For the parameters of the figure, the caustics do not map regions of particularly large $\hs$. This is unsurprising since $\gamma$ is at most $0.447$. 

To assess how large $\gamma$ or equivalently $s$ need to be for caustics to be the dominant feature of $\Hs$, we show in figure \ref{fig:caustics1} $\hs$ computed from MA for $U_m = 0.8$ m s$^{-1}$ and $s = 200$ (left panel, $\gamma = 1$) and $s=4000$ (right panel, $\gamma = 4.47$). It is only for $s=4000$ that the caustics are evidently controlling the significant wave height pattern. We emphasise that $s=200$ and a fortiori  $s=4000$ are unrealistically large values: observational estimates for $s$ in the open ocean seldom exceed $s=80$. We conclude that caustics are unlikely to play a role in real ocean conditions.

\begin{figure}
 \includegraphics[width=\textwidth]{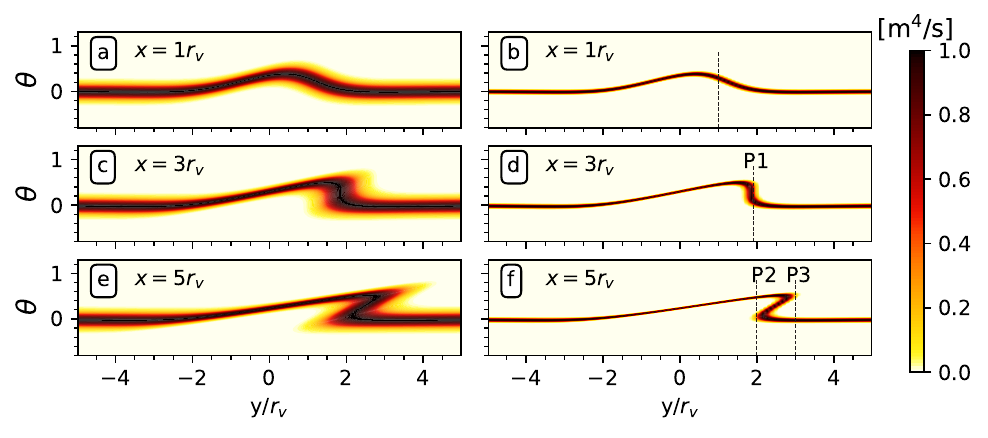}
    \caption{Wavenumber-integrated action density $\int \mathcal{A}(x,y,k,\theta) \dd k$ as a function of $y$ and $\theta$ for $x=r_v, 3 \, r_v$ and $5 \, r_v$ corresponding to the significant wave height shown in figure \ref{fig:caustics1} for $s=200$ (left column) and $s=4000$ (right column). P1 in panel d corresponds to the values of $(x,y)$  of the cusp  from where the caustics emanate; P2 and P3 are associated to points on each of the two caustics. }
\label{fig:caustics2}
\end{figure}

With academic rather than practical interest in mind, then, we show in figure \ref{fig:caustics2} the integrated action $\int \Ac \, \dd k$ as a function of $y$ for different three different values of $x$ (identified by dashed vertical lines in figure \ref{fig:caustics1}). The figure illustrates how caustics emerge from a fold singularity in the surface $\mathcal{S}(x,y,\theta)=0$ along which action is concentrated in the $(x,y,\theta)$ phase space. For $x = r_v$, the surface is a graph over $(x,y)$ and there are no caustics; for $x = x_c \approx 3 r_v$, the surface has a single point of vertical tangency (P1 in panel (f) of \ref{fig:caustics1}) corresponding to the birth of caustics at a cusp in the $(x,y)$-plane; for $x = 5 r_c$, there are two points of vertical tangency, P2 and P3 in panel (h), corresponding to the two caustic curves. The picture is increasingly blurred as $s$ decreases (compare the right panel of figure \ref{fig:caustics2} with the left panels and with figure \ref{A_comparison}), explaining the diminishing importance of  caustics for $\Hs$.

\section{Discussion and conclusion}\label{sec:discussion}

The main  results in this study   are obtained by approximate solution of the wave action equation in the four-dimensional position--wavenumber space. \change{A main organizing principle identified by the analysis  is that scattering of SGWs by spatially compact currents results in the deflection function, $\Delta(y)$ in \eqref{inner7}. Although $\Delta$ varies linearly with the vertical vorticity of the currents, $\Delta$  figures in a nonlinear transformation of the action density.  This nonlinear transformation produces the  modulation of the significant wave height $\Hs$ behind the scattering region, e.g. the expression for $\Hs$ in \eqref{HsSumm}. Quantities that depend on other moments (e.g., Stokes drift) behave similarly and could be readily inferred from our explicit forms \eqref{inner3} and \eqref{AcXO1-1} for the wave action density.}

\change{While we have obtained these results for deep-water SGWs, they apply essentially unchanged to other two-dimensional waves with isotropic dispersion relation such as finite-depth SGWs or Poincar\'e waves. The conclusions we draw  about $\Hs$ can also be rephrased in terms of other root-mean-square quantities relevant to waves other than SGWs. With a little effort, the approach we adopt, based on the matched asymptotics treatment of the wave action equation, could be further extended to three-dimensional waves and to anisotropic dispersion relations. Our results could easily be extended to account for vertically sheared currents using the modified dispersion relation of \citet{kirby1989surface} (which involves a Doppler shift term that is nonlinear in $\bk$).}

In addition to the WKB approximation used to derive the action conservation equation \eqref{actioncons} there are two independent  approximations involved:
\begin{enumerate}
\item[(a)] the current speed is much less than the group velocity of the incident swell;
\item[(b)] swell with small directional spreading is incident on a region of spatially compact currents e.g. an axisymmetric vortex or a vortex dipole.
\end{enumerate}
Provided that (a) and (b) are satisfied the approximate  solution of the wave action equation compares well with numerical solutions provided by WW3.

Approximation  (a) is usually justified. To challenge   (a) one must   consider   current speeds such as 2 m s$^{-1}$ e.g. observed as  a  peak current speed in the Agulhas  system \citep{Quilfen2019}.  Swell with 100 m wavelength has  group velocity  $\sim 6$ m s$^{-1}$ so that the small parameter in (a)  is as large as  $1/3$. In less extreme cases  approximation (a) will be satisfied. 

Approximation (b) is less secure: ocean swell is not sufficiently  unidirectional to strongly justify (b) e.g. see the $\delta$-column in table \ref{tb:parameters}.  Over long distances, the continuous scattering by uncorrelated currents leads to a broadening of the angular spectrum. When approximation (a) applies, this broadening is described by the directional diffusion equation for wave action derived by \citet{VillasBoas2020b}. 
This diffusion process is one of the mechanisms that makes swell with very small values of $\delta$ unlikely. \change{However, our computations for a Gaussian vortex indicate that our asymptotic results 
are reliable for the moderately small values of $\delta$ typical of swell.} 

Because of the relatively large directional spreading of ocean swell the mathematical ideal of a sharp  wave caustic is not realized. Instead  the caustic singularity is `washed out' \citep{Heller2008}. Behind a vortex we find instead an elongated streaky  pattern in $\Hs$.

\change{Our results show that $\Hs$ behind an axisymmetric vortex with parameters in table \ref{tb:parameters} has spatial variation as large as $\pm 30\%$ of the incident constant value $\Hss$. Spatial inhomogeneities in $\Hs$ of this magnitude are important for wave breaking and exchange of  momentum, heat and gas between the ocean and atmosphere. For example,  airborne observations of the ocean surface  by \cite{Romero2017} indicate that $\pm 30\%$ variations in  $\Hs$ are associated with an order of magnitude increase in whitecap coverage.}

The directional diffusion equation  of \cite{VillasBoas2020b} uses only approximation (a). One does not need to assume  that the wave field is strongly unidirectional or that the currents are spatially compact. Moreover the directional diffusion equation is obtained without detailed consideration of the   perturbations to the action spectrum that accompany wave scattering. But there is useful information hiding  in these unexamined   perturbations to the action spectrum. We are currently engaged in extracting these perturbations, calculating the attendant spatial variability to $\Hs$,  and relating the statistics of these fluctuations in $\Hs$ to those of the surface currents. These future developments promise to explain numerical experiments that  identify   relations between the spectral slopes of surface-current spectra and those of significant wave height \citep{VillasBoas2020}.

\backsection[Acknowledgements] {We thank Victor Shrira for conversations about this work.}

\backsection[Funding]{JV and HW are  supported by the UK Natural Environment Research Council (grant NE/W002876/1). ABVB is supported by NASA award 80NSSC23K0979 through the International Ocean Vector Winds Science Team. WRY  is supported by the National Science Foundation award 2048583.}

\backsection[Declaration of interests]{The authors report no conflict of interest.}

\backsection[Data availability statement]{The WW3 configuration files applied in this work can be found at \url{https://www.cambridge.org/S0022112023006869/JFM-Notebooks/files/Figure-3.ipynb}. The \change{customizable} Jupyter Notebook file demonstrating the matched asymptotics approach is available at \url{https://www.cambridge.org/S0022112023006869/JFM-Notebooks/files/Figure-3.ipynb}.}

\backsection[Author ORCID]{H. Wang, https://orcid.org/0000-0002-5841-5474. A. B. Villas B\^{o}as, https://orcid.org/0000-0001-6767-6556.  W.R Young,  https://orcid.org/0000-0002-1842-3197.  J. Vanneste, https://orcid.org/0000-0002-0319-589X.}

\appendix

\section{Incident spectrum \label{modelAppen}}

We use the separable spectrum
\beq
\Fcs(k,\theta)= \zetrmss^2F_\star(k)\, \Ds(\theta) . 
\label{LHCS3}
\eeq  
The wavenumber function in   \eqref{LHCS3} is 
\beq
F_\star(k)\defn \frac{2}{\erfc(-\sigmas/\sqrt{2} \delsig)} \frac{\ee^{ -(\sigma - \sigmas)^2/2 \delsig^2}}{\sqrt{2 \pi \delsig^2} \,} \, \frac{1}{k} \frac{\dd \sigma}{\dd k}, 
\label{LHCS37}
\eeq
where $\erfc$ is the complementary error function.  It corresponds to a Gaussian spectrum in frequency truncated at $\sigma=0$.
The  angular part of the spectrum in \eqref{LHCS3} is
\beq
\Ds(\theta) \defn \frac{\Gamma(s+1)}{2\sqrt{\pi} \Gamma(s+\half)} \, \cos^{2s} \left(\frac{\theta}{2}\right)
\label{LHCS43}
\eeq
\citep[][]{LHCS1963}, which corresponds to incoming waves spread around $\theta=0$.
The four parameters in this model spectrum are the root mean square sea-surface displacement $\zetrmss$, the peak radian frequency  $\sigmas=\sqrt{g \ks}$, the spectral width $\delsig$ and the directional spreading parameter $s$.
Normalization is ensured with 
\beq	
\int_{-\pi}^\pi \!\!\!\Ds(\theta) \, \dd \theta = 1 \qquad \text{and} \qquad \int_0^\infty\!\!\!\! F_\star(k) k \, \dd k =1.
\label{LHCS47}
\eeq

In the narrow-band limit $\delsig/\sigmas \ll 1$ and $s \gg 1$, the spectrum is approximated by \eqref{LHCS54}
with $\delta_k = 2 \delta_\sigma \sqrt{\ks/g}$ and $\delta_{\theta}=\sqrt{2/s}$. The parameter $\delta_{\theta}$ captures the standard deviation in the angular distribution, which is the definition of `directional spreading' \citep{kuik1988method}. We note that the expressions for directional spreading are sometimes formally different, but equivalent to our expression for  $\delta_{\theta}$ at large $s$. For example, another popular way to state the definition for a generic directional distribution is  
\beq \label{dspreadWW}
\sigma_{\theta} \defn \LL[2\LL(1-\LL(a^2+b^2\RR)^{1/2}\RR)\RR]^{1/2},
\eeq
where
\beq
a=\int\cos{\theta}\Ds(\theta)\, \dd \theta \quad \textrm{and} \quad b=\int\sin{\theta}\Ds(\theta)\, \dd \theta \label{ab}
\eeq
\citep{VillasBoas2020}.
Using the expression of $\Ds$ in \eqref{LHCS54}, we can compute the integrals in \eqref{ab} analytically, getting $a=\ee^{-1/s}$ and $b=0$. Therefore, 
\beq
\sigma_{\theta}^2=2(1-\ee^{-1/s}) \to 2/s \quad \textrm{as} \ \ s\to \infty.
\eeq
Thus the definition of $\sigma_{\theta}$  in \eqref{dspreadWW} indeed agrees with the parameter $\delta_{\theta}$ at large $s$.

\section{Set up of WAVEWATCH III} \label{app:WW3}

We compare our results with numerical simulations from an idealized setup of WW3 which integrates the action balance equation \eqref{actioncons}. Here, we focus on freely propagating swell-type waves, so the effects of wind forcing, nonlinear interactions and wave breaking are ignored \citep[e.g.,][]{VillasBoas2020}. We use WW3 version v6.07.1 (\url{https://github.com/NOAA-EMC/WW3/releases/tag/6.07.1}) to solve \eqref{actioncons} on a 1000~km$\times$1000~km Cartesian domain with 5~km grid spacing. To resolve  swells with $s=10$ and $40$ the spectral grid has 80 directions and 32 frequencies.  Larger values of $s$ (i.e., narrower directional spreading) would require higher directional resolution for the model to converge. We use a global integration time step of 200~s, spatial advection time step of 50~s, spectral advection time step of 12~s, and minimum source term time step of 5~s. We verified that decreasing the time stepping  or the spatial grid spacing does not significantly change the results (not shown).

All simulations are initialized with the narrow-banded wave spectrum in \eqref{LHCS3}. Waves enter the domain from the left boundary with initial mean direction $\theta = 0^\circ$ (propagating from left to right), directional spreading parameter $s=10$ or $s=10,$ peak frequency $\sigmas = 0.61$ rad s$^{-1}$ (peak period of 10.3 s), spectral width $\delta_{\sigma} = 0.04$,  and $\Hss$ = 1 m. The boundary condition at the left boundary is kept constant throughout the experiment and each experiment is run until steady state is reached.

As mentioned in $\S$\ref{sec:gaussian}, a control run is conducted in the absence of currents. Although there is no scattering from the currents, a nonuniform $\hs^{\mathrm{ctrl}}=\Hs^{\mathrm{ctrl}}-\Hss$ arises, due to the limited domain size in $y$, which leads to a reduction of incident wave action from waves arriving 
from large $|y|$ --- an effect that is more pronounced at large $x$. 
As $s$ increases, the action density in the incident spectrum is more concentrated in the eastward direction, leading to less leakage of wave action through the top and bottom boundaries and a more spatially uniform $\hs^{\mathrm{ctrl}}$. This leakage of wave action corresponds to a reduction of 5\% in $\hs^{\mathrm{ctrl}}$ for $s=10$, and 2\% for $s=40$ towards the right-hand side boundary.

\section{MA--WW3 mismatch in the scattering region} \label{app:mismatch}

\begin{figure}
  \centering
  \includegraphics[width=1.0\textwidth]{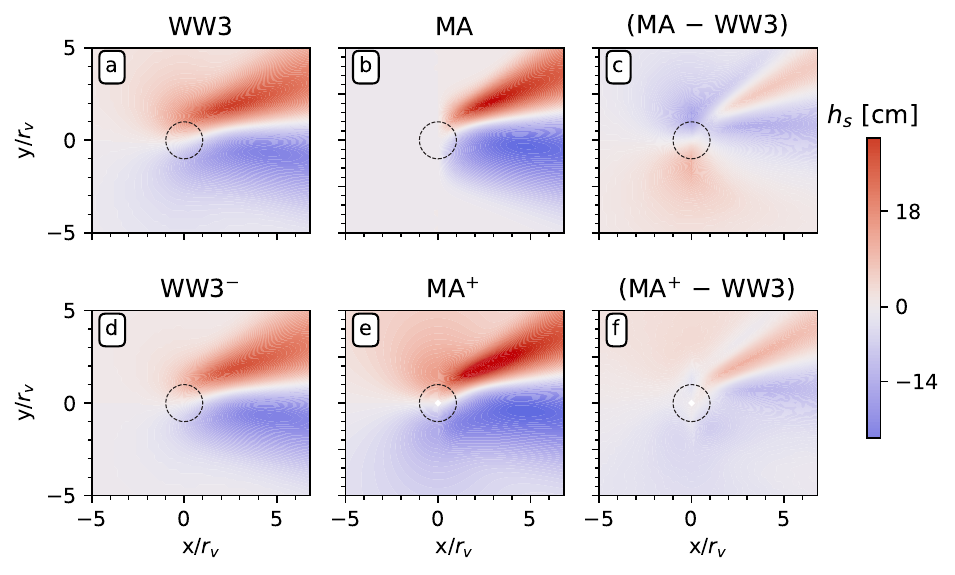}
    \caption{Significant wave height anomaly  $\hs$  computed from (a) WW3, (b) MA (same as figures 3g,h), (d) WW3$^{-}$ (where the term proportional to $\partial_k \Ac$ is switched off) and (e)  MA$^{+}$as in (C6). Panels (c) and (f) show the differences between (a) and (b), and (a) and (e), respectively. All plots have the same colour bar.
}
  \label{GV_concertina}
\end{figure}

We develop a heuristic correction to MA that we show captures the non-zero $\hs$ in the scattering region. First, we note that the non-zero $\hs$ in the scattering region from WW3 appears localized, likely caused by the term proportional to $\partial_k\Ac$  in \eqref{pkA}, as the terms proportional to $\partial_{\theta}\Ac$ result in non-local effects. This observation is confirmed by a WW3 run, which we refer to as WW3$^{-}$, where the term in $\partial_k \Ac$ is suppressed in the wave action equation, yielding a more uniform  $\hs$ in the scattering region (see panel (d) in Figure \ref{GV_concertina}). We then recall  that in the MA solution, the insignificance of the $\partial_k\Ac$ term is due to the approximation of a single dominant wavenumber in the steps leading to \eqref{HsSc}. We thus return to the approximation \eqref{xO1} of the wave-action transport equation in the scattering region and relax the approximation of replacing $k$ by $\ks$. 
We focus on the $\theta$-integrated action
\beq
\Bc(\bx,k) = \int \Ac(\bx, \bk) \, \dd \theta.
\eeq
It satisfies
\beq
c(k) \,  \partial_x \Bc - U_x(\bx) k  \,\partial_k \Bc = 0.
\eeq
Noting that $c(k) = g^{1/2} k^{-1/2}/2$, we solve this equation using the method of characteristics to find
\beq
\Bc(\bx,k) = \Bc_\star\left( \left(k^{-1/2} - g^{-1/2} U(\bx) \right)^{-2}  \right).
\eeq
The significant wave height is deduced by integration as
\beq
\Hs(\bx) = \left( \frac{16}{g^{1/2}} \int \Bc_\star\left( \left(k^{-1/2} - g^{-1/2} U(\bx) \right)^{-2}  \right) k^{3/2} \, \dd k \right)^{1/2}.
\eeq
We now change the integration variable, taking advantage of the localisation of $\Bc_\star(k)$ to ignore the corresponding change in the lower limit of integration and obtain
\begin{align}
 \Hs(\bx) &= \left( \frac{16}{g^{1/2}} \int \Bc_\star(k) \left(k^{-1/2} + g^{-1/2} U(\bx) \right)^{-6} k^{-3/2} \, \dd k \right)^{1/2} \nonumber \\
 &= \left( \frac{16}{g^{1/2}} \int \Bc_\star(k) k^{3/2} \left(1 + k^{1/2} g^{-1/2} U(\bx) \right)^{-6} \, \dd k \right)^{1/2} 
\nonumber \\
&= \left( \frac{16}{g^{1/2}} \int \Bc_\star(k) k^{3/2} \left(1 +  \frac{U(\bx)}{2c(k)} \right)^{-6} \, \dd k \right)^{1/2}
\end{align}
At this point, we can approximate $c(k)$ by $c_\star$ in the small, $O(\eps)$ term $U(\bx)/(2c(k))$ and use two binomial expansions to obtain
\beq
\Hs(\bx) \approx \Hss \left(1 \change{-} \frac{3 U(\bx)}{2 \cs} \right). \label{MAplus}
\eeq 
We emphasise the heuristic nature of this approximation (MA$^+$) which is formally no more accurate than the MA approximation $\Hs(\bx)=\Hss$ since it neglects some, though not all, $O(\delta)$ terms. Nonetheless, it captures most of the significant wave height anomaly close to the Gaussian vortex, as figure \ref{GV_concertina} demonstrates under parameters $s=40$ and $U_m=0.8$ m s$^{-1}$.

\section{Caustics for the Gaussian vortex} \label{app:caustics}

In the Gaussian vortex example, we can derive the locations of the caustics in the $(x,y)$ plane analytically. Using expression \eqref{Del77}  for $\Delta(y)$ and introducing the functions 
\beq \label{chanvar1}
w(x,y) \defn -(y-x\theta)^2/r_v^2
\eeq 
and
\beq \label{chanvar2}
q(x) \defn-2\pi r_v^4 \cs^2/(x^2 \kappa^2),
\eeq
we can write equations \eqref{cloc} defining the caustics as 
\beq\label{ceq1}
\theta-\frac{\kappa}{\sqrt{2\pi}r_v\cs}\ee^{w/2}=0 
\eeq
and
\beq\label{lambert}
\qquad w\ee^{w}=q.
\eeq
Eq.\ \eqref{lambert} relates $w$ to $q$, and takes the standard form defining the Lambert $W$-functions \citep[see][Eq.~4.13.1]{DLMF}.
This equation has two branches of solutions $w=W_{i}(q)$, $i=0, \, -1$,  when $0  < -q < \ee$ and no solutions when $-q > \ee$ ($q < 0$ by definition \eqref{chanvar2}).   
The two branches meet at $q=-\ee^{-1}$ which corresponds to 
\beq
x=x_c\defn \sqrt{2\pi e}r_v^2\cs/\kappa. \label{xc}
\eeq 

Physically, the two branches $w=W_{i}(q)$ correspond to two caustic lines in the $(x,y)$ plane that emanate from a cusp  point with $x=x_c$.
The equation of the caustics is found using \eqref{chanvar1} and \eqref{ceq1} as
\beq \label{causyx}
y=\frac{\kappa x\,  \ee^{W_{i}(q(x))/2}}{\sqrt{2\pi}r_v\cs}+\sqrt{-W_{i}(q(x))}r_v,\quad x\geq x_c.
\eeq
The cusp point  is at $(x,y)=(x_c,2 r_v)$. 

The asymptotic form of the caustics for $x \to \infty$ is readily obtained by noting that $q(x) \to 0^-$ as $x \to \infty$ and then that $W_0(q) \to 0$ and $W_{-1}(q) \sim \ln (-q)$. Thus the $i=0$ caustic asymptotes to a straight line and the $i=-1$ caustic to $y \sim (2 \ln x)^{1/2}$.

\bibliographystyle{jfm}
\bibliography{swellcurrents_productionfile2}

\end{document}